\documentclass[prd]{revtex4}

\usepackage{epsfig}

\def\etal{{\it et al.}}

\begin{document}


\title{Charmonium in China: BEPCII/BESIII}

\author{Frederick A. Harris 
For the BES Collaboration}
\affiliation{Department of Physics and Astronomy, \\
        The University of Hawaii, \\
        Honolulu, Hawaii 96822, USA \\
fah@phys.hawaii.edu}

\begin{abstract}
  Results based on 106 M $\psi'$ events and about 226 M $J/\psi$
  events acquired with the BESIII detector at BEPCII are reported,
  including a confirmation of the BESII $p\bar{p}$ threshold
  enhancement in $J/\psi \to \gamma p\bar{p}$, branching ratios for
  $\chi_{cJ} \to \pi^0 \pi^0$ and $\eta \eta$, and first measurements
  of the branching ratios for $\psi' \to \pi^0 h_c$ and $h_c \to
  \gamma \eta_c$, as well as measurements of the mass and width of the
  $h_c$. Also reported are first observations of the two photon
  process $\psi' \to \gamma \gamma J/\psi$ and $f_0(980) - a_0(980)$
  mixing.
 
\end{abstract}

\maketitle


\section{Introduction}  
The Beijing Electron-Positron Collider has been upgraded to a two-ring
collider (BEPCII) with a design luminosity of $1 \times
10^{33}$cm$^{-2}$ s$^{-1}$ at a center-of-mass energy of 3.78 GeV.
It will operate between 2 and 4.6 GeV, allowing precision
studies of charmonium ($J/\psi$, $\psi'$, $\psi(3770)$, $\eta_c$,
$\chi_{cJ}$, and $h_c$), charm ($D$ and $D_s$ mesons), and improved
determinations of the tau mass and the hadronic cross section ($R$) in
this energy region. Details on BESIII and BEPCII are described in
Ref.~\cite{fah_meson08}.

During 2009, BESIII acquired a sample of 106 M $\psi'$ events, or four
times the CLEOc sample, and about 226 M $J/\psi$ events, or about four
times the BESII $J/\psi$ sample.  Results in this paper are based on
these data sets.

\section{\boldmath $h_c$ studies}

In 2005, CLEO~\cite{cleo} reported a measurement of the mass of the
$h_c(^1P_1)$ in $e^+ e^- \to \psi' \to \pi^0 h_c,$ $h_c \to \gamma
\eta_c$, where they used both inclusive and exclusive $ \eta_c$ decay
events.  In 2008, they repeated their analysis with 25 M $\psi'$
events.~\cite{cleohc}
Combining results, they obtain $m(h_c)_{AVG} = 3525.2 \pm 0.18 \pm
0.12$ MeV/c$^2$.~\cite{cleohc} A precise determination of the mass is
important to learn about the hyperfine (spin-spin) interaction of
$P$ wave states.  Using the spin weighted centroid of the $^3P_J$
states, $<m(^3P_J)>$, to represent $m(^3P_J)$, they obtain $\Delta
m_{hf}(1P) =~<m(^3P_J)> - m(^1P_1) = +0.08 \pm 0.18 \pm 0.12$
MeV/c$^2$.  This is consistent with the lowest order expectation of
zero.

BESIII has measured $m(h_c)$ from the distribution of mass recoiling
against $\pi^0$s, both in inclusive $\psi' \to \pi^0 h_c$ samples and
with samples tagged by the $E1$ photon from $h_c \to \gamma \eta_c$.
This allows for the first time determinations of $B(\psi' \to \pi^0
h_c)$ and $B(h_c \to \gamma \eta_c)$. BESIII also measures for the
first time the width $\Gamma(h_c)$.  Results are shown in
Table~\ref{hc} and compared with CLEOc and theory. More
detail may be found in Ref.~\cite{bes3hc}.

\begin{table}[htb]
\caption{BESIII $h_c$ results compared with CLEOc and theory.}
{\begin{tabular}{@{}lccc@{}} \toprule
     & BESIII & CLEOc~\cite{cleohc} & Theory \\ \colrule
$B(\psi' \to \pi^0 h_c)$
& $4.58 \pm 0.40 \pm 0.50$ & $4.16\pm0.30\pm0.37$ \\ 
$\times B(h_c \to \gamma \eta_c)$ [$10^{-4}$]\\ \colrule
$B(\psi' \to \pi^0 h_c)$ [$10^{-4}$]
& $8.4 \pm 1.3 \pm 1.0$ & & $(4 - 13)$~\cite{kuang}\\ \colrule
$B(h_c \to \gamma \eta_c)$ [\%]
& $54.3\pm6.7\pm5.2$ & & 41 (NRQCD)~\cite{kuang} \\ 
 & & & 88 (PQCD)~\cite{kuang} \\ 
 & & & 38~\cite{rosner}  \\ \colrule
$m(h_c)$ [MeV/c$^2$] & $3525.40\pm0.13\pm0.18$ &
$3525.20\pm0.18\pm0.12$ \\ \colrule
$\Gamma(h_c)$ [MeV/c$^2$] & $0.73 \pm0.45\pm0.28$ & & 1.1 (NRQCD)~\cite{kuang} \\
                          & $< 1.44$ @ 90\% C.L.  & & 0.51 (PQCD)~\cite{kuang}  \\ \colrule
$\Delta M_{hf}(1P)$ [MeV/c$^2]$ & $0.10 \pm0.13\pm0.18$ &
$0.08\pm0.18\pm0.12$ \\ \colrule
\end{tabular} \label{hc}}
\end{table}

\section{\boldmath $\chi_{cJ}$ decays}

Because of the large $\psi'$ sample and the large branching fractions
for $\psi' \to \gamma \chi_{cJ}$, BESIII is in a good position to
study both inclusive and exclusive $\chi_{cJ}$ decays.  Since
$\chi_{cJ}$ decays to two gluons, these decays are a good place to
search for gluonium.~\cite{amsler} The color octet mechanism is
important in these decays,~\cite{bolz} and measurements allow tests of
theoretical models.

BESIII has studied $\psi' \to \gamma \chi_{cJ}, \chi_{cJ} \to \pi^0
\pi^0$ and $\eta \eta$, where $\pi^0$ and $\eta$ decay to $\gamma
\gamma$.~\cite{bes3pi0pi0} Results are given in Table~\ref{pi0pi0} and
compared to CLEOc~\cite{CLEOcpi0pi0}, the particle data tables
(PDG08),~\cite{PDG08} and theory~\cite{bolz}.  Improved measurements
will allow refinement of theory.

\begin{table}[htb]
\caption{Branching fraction results for $\chi_{cJ} \to \pi^0 \pi^0$ and
  $\eta \eta$. The last errors for BESIII and CLEOc are the branching
  fraction uncertainties for $\psi' \to \gamma \chi_{cJ}$. Note that
  CLEOc used their own $\psi' \to \gamma \chi_{cJ}$ branching
  fractions for their results.}
{\begin{tabular}{@{}lccc@{}} \toprule
Decay  &  & $\chi_{c0}$ $(10^{-3})$  & $\chi_{c2}$ $(10^{-3})$  \\ \colrule
$\pi^0\pi^0$ & BESIII~\cite{bes3pi0pi0} & $3.23\pm0.03\pm0.23\pm0.14$ & $0.88\pm0.02\pm0.06\pm0.04$ \\
 & CLEOc~\cite{CLEOcpi0pi0} & $2.94\pm0.07\pm0.32\pm0.15$ & $0.68\pm0.03\pm0.07\pm0.04$ \\
 & PDG08~\cite{PDG08} & $2.43\pm0.20$ & $0.71\pm0.08$ \\ 
 & Theory~\cite{bolz} & 2.3  & 0.95 \\ \colrule
$\eta\eta$ & BESIII~\cite{bes3pi0pi0} & $3.44\pm0.10\pm0.24\pm0.20$ & $0.65\pm0.04\pm0.05\pm0.03$ \\
 & CLEOc~\cite{CLEOcpi0pi0} & $3.18\pm0.13\pm0.31\pm0.16$ & $0.51\pm0.05\pm0.05\pm0.03$ \\ 
 & PDG08~\cite{PDG08} & $2.4\pm0.4$ & $<0.5$ \\ 
 & Theory~\cite{bolz} & $3.2$ & $1.3$ \\ \botrule

\end{tabular} \label{pi0pi0}}
\end{table}

\begin{figure}[!htb]
\centerline{\psfig{file=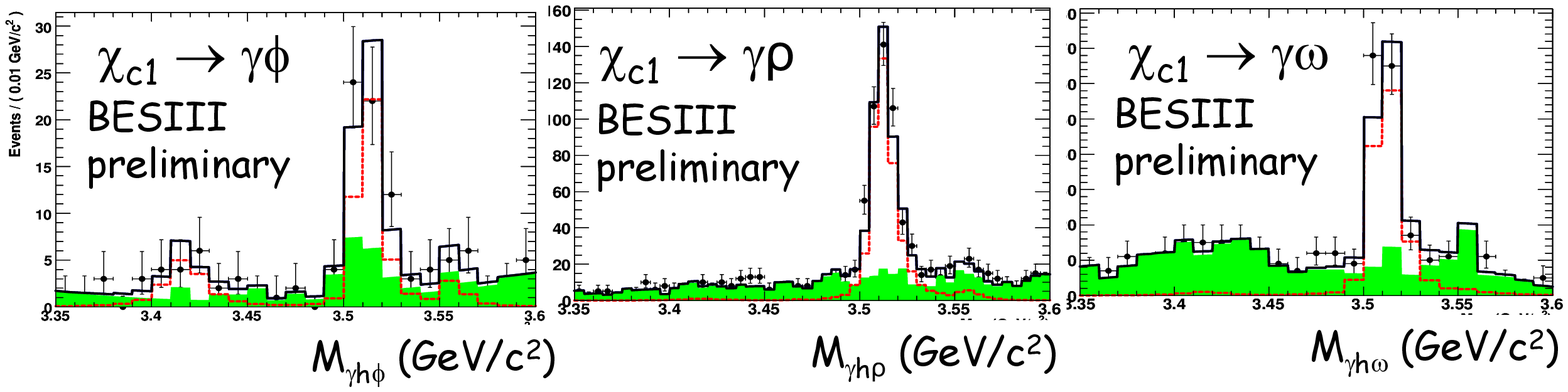,width=12.0cm}}
\vspace*{8pt}
\caption{Invariant mass distributions of the high energy gamma and
  $V$ from $\psi' \to \gamma \chi_{cJ}, \chi_{cJ} \to \gamma V$, where
  $V$ is $\phi$, $\rho$, or $\omega$. Dots with error bars are data,
  the shaded histogram is background estimated from Monte Carlo
  simulation, and the open black histogram is the fit. Clear signals
  for $\chi_{c1} \to \gamma V$ are observed.
  \label{fig:gammaV}}
\end{figure}

BESIII is studying $\psi' \to \gamma \chi_{cJ}, \chi_{cJ} \to \gamma
V$, where $V$ is $\phi \to K^+ K^-$, $\rho \to \pi^+ \pi^-$, or
$\omega \to \pi^+ \pi^- \pi^0$. Invariant mass distributions of the
high energy gamma and $V$ are shown in Fig.~\ref{fig:gammaV}, where
clear signals for $\chi_{c1} \to \gamma V$ are observed.  Preliminary
results are given in Table~\ref{table:gammaV} and compared to
CLEOc~\cite{CLEOcgV} and perturbative QCD predictions,~\cite{pQCD} which
for $\chi_{c1} \to \gamma V$ are an order of magnitude too low.
BESIII observes $\chi_{cJ} \to \gamma \phi$ for the first time.

\begin{table}[!htb]
\caption{Branching fraction results for $\chi_{cJ} \to \gamma V$, where
  $V$ is a $\phi$, $\rho$, or $\omega$. For BESIII, results are
  preliminary, and errors are
  statistical only.}
{\begin{tabular}{@{}lccc@{}} \toprule
Decay  & BESIII $(10^{-6})$ & CLEOc~\cite{CLEOcgV} $(10^{-6})$  & pQCD~\cite{pQCD} $(10^{-6})$  \\ \colrule
$\chi_{c0} \to \gamma \phi$ & $<14.8$ & $<6.4$ & 0.46 \\ 
$\chi_{c1} \to \gamma \phi$ & $27.3\pm5.5$ & $<26$ & 3.6 \\ 
$\chi_{c2} \to \gamma \phi$ & $<7.8$ & $<13$ & 1.1 \\ \colrule
$\chi_{c0} \to \gamma \rho^0$ & $<9.5$ & $<9.6$ & 1.2 \\ 
$\chi_{c1} \to \gamma \rho^0$ & $241\pm14$ & $243\pm19\pm22$ & 14 \\ 
$\chi_{c2} \to \gamma \rho^0$ & $<19.7$ & $<50$ & 4.4 \\ \colrule
$\chi_{c0} \to \gamma \omega$ & $<11.7$ & $<8.8$ & 0.13 \\ 
$\chi_{c1} \to \gamma \omega$ & $73.5\pm7.6$ & $83\pm15\pm12$ & 1.6 \\ 
$\chi_{c2} \to \gamma \omega$ & $<5.8$ & $<7.0$ & 0.5 \\ \botrule
\end{tabular} \label{table:gammaV}}
\end{table}

BESIII has also fit the helicity angle distributions in the decays
$\chi_{c1} \to \gamma V$.  The helicity angle is the angle between the
$V$ direction in the $\chi_{c1}$ rest frame and a daughter meson in
the vector meson rest frame ($\rho$ or $\phi$) or the normal to the
decay plane in the $\omega$ rest frame.  Longitudinal (transverse)
polarization exhibits a $\cos^2 \theta$ ($\sin^2 \theta$)
dependence.
Results for $f_t$, where 
\begin{equation}
f_t = \frac{N_t}{N_t + R*N_l},
\end{equation}
$N_t$ is the number of fitted transversely polarized events, $N_l$ is the
number of fitted longitudinally polarized events, and $R$ is the ratio of the
efficiencies for transversely and longitudinally polarized events, are
given in Table~\ref{table:ft}, along with results from
CLEOc.~\cite{CLEOcft} The helicity distributions indicate that the
vector mesons are preferentially longitudinally polarized.
  

\begin{table}[htb]
\caption{Results of fits to helicity angle distributions in $\chi_{c1} \to
  \gamma V$ . BESIII results are preliminary.}
{\begin{tabular}{@{}lcc@{}} \toprule
Decay     & $f_t$-BESIII & $f_t$-CLEOc~\cite{CLEOcft} \\ \colrule
$\chi_{c1} \to \gamma \rho$ & $0.155\pm0.033\pm0.014$ &
$0.072^{+0.046+0.002}_{-0.035-0.022}$ \\ 
$\chi_{c1} \to \gamma \omega$ & $0.240^{+0.091+0.044}_{-0.086-0.027}$ &
$0.32^{+0.27+0.10}_{-0.19-0.19}$ \\ 
$\chi_{c1} \to \gamma \phi$ & $0.27^{+0.13+0.08}_{-0.12-0.08}$ & \\ \botrule
\end{tabular} \label{table:ft}}
\end{table}

\section{\boldmath $\psi' \to \gamma \gamma J/\psi$}
\begin{figure}[!htb]
\centerline{\psfig{file=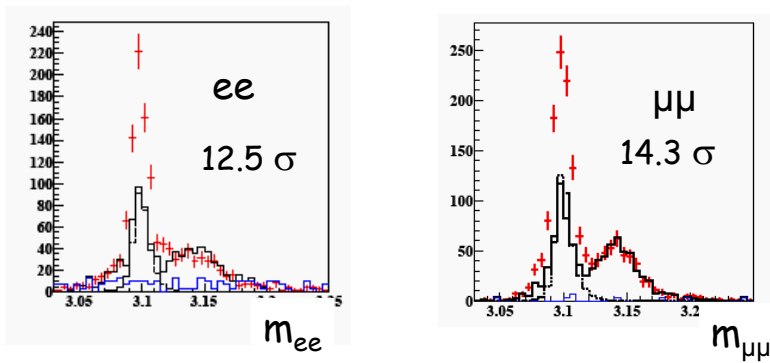,width=10.0cm}}
\vspace*{8pt}
\caption{Invariant mass distributions of $e^+e^-$ and $\mu^+ \mu^-$ in
  $\psi' \to \gamma \gamma J/\psi, J/\psi \to l^+l^-$ events. Dots
  with error bars are data.  The solid black histogram is the sum of
  $\psi'$ background processes from Monte Carlo simulation and
  background from continuum.  A clear excess is seen at the $J/\psi$ peak.  
  \label{2gamma}}
\end{figure}

Two photon transitions are well known in excitations of molecules,
atomic hydrogen, and positronium but except for the discrete
transitions $\psi' \to \gamma \chi_{cJ}, \chi_{cJ} \to \gamma J/\psi$
have never been observed in charmonium systems.  BESIII has searched
for continuous two photon transitions using $\psi' \to \gamma \gamma
J/\psi, J/\psi \to l^+ l^-$ events, where $l$ is $e$ or $\mu$.
Selection requirements are used to reduce backgrounds from $\psi' \to
\gamma \chi_{cJ}, \chi_{cJ} \to \gamma J/\psi$ and $\psi' \to \pi^0
J/\psi$ and $\eta J/\psi$.  The invariant mass distributions of the
two leptons for selected events are shown in Fig.~\ref{2gamma}.  A
clear excess is seen at the $J/\psi$ peak over $\psi'$ background from
Monte Carlo simulation plus continuum background. Branching fractions
of $B(\psi' \to \gamma \gamma J/\psi)$ of $(1.06\pm0.08)\times
10^{-3}$ and $(0.99\pm0.07)\times 10^{-3}$ are determined for the
$J/\psi \to ee$ and $J/\psi \to \mu \mu$ cases, respectively.
Combining, we determine $B(\psi' \to \gamma \gamma J/\psi) =
(1.02\pm0.05^{+0.19}_{-0.20})\times 10^{-3}$.  This is a first
measurement. The result is preliminary, and more details can be found
in Ref.~\cite{xiaorui}.

\section{\boldmath $f_0(980) - a_0(980)$ mixing}

\begin{figure}[!htb]
\centerline{\psfig{file=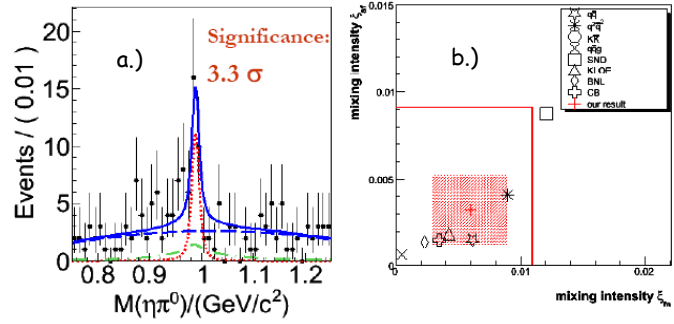,width=9.0cm}}
\vspace*{8pt}
\caption{a.) Invariant mass of $\eta\pi^0$ in $J/\psi \to \phi f_0 \to
  \phi a_0 \to \phi \eta \pi^0$.  The mixing signal is very narrow
  ($<8$ MeV/c$^2$). b.) Mixing intensity $\xi_{fa}$ versus $\xi_{af}$
  for various models and calculations and our results (shaded square)
  and upper limits (lines). 
  \label{fig:f0mixing}}
\end{figure}

The $f_0(980)$ and $a_0(980)$ are controversial particles.  They have
been described as $q\bar{q}$ or $q\bar{q}q\bar{q}$ states, $K \bar{K}$
molecules, or $q\bar{q}G$ hybrid states. Mixing was first suggested by
Achasov,~\cite{achasov} and mixing measurements may be important to
clarify the nature of these particles. There have been
suggestions~\cite{f0mixing} for BESIII to search for mixing signals in
$J/\psi \to \phi f_0 \to \phi a_0 \to \phi \eta \pi^0$, where the
$f_0$ mixes to $a_0$, and in the process~\cite{a0mixing} $\chi_{c1} \to
\pi^0 a_0 \to \pi^0 f_0 \to \pi^+\pi^-\pi^0$, where $a_0$ mixes to
$f_0$. The signal for the former case is shown in
Fig.~\ref{fig:f0mixing} a).  The expected mixing signal is very narrow
(8 MeV/$^2$) between the $K^+K^-$ and $K^0_S K^0_S$ thresholds (987 -
995 MeV/$^2$).  Backgrounds are sideband and a wide $a_0$ from $J/\psi
\to \gamma^*/K^*K \to \phi a_0$. Fitting to signal plus backgrounds
determines $24.7\pm8.6$ mixing events or $< 36.7 $ at the 90\%
confidence level (C.L.).  This gives a mixing intensity $\xi_{fa} =
(0.6\pm0.2\pm0.2)\%$ or $<1.1 \%$ at the 90\% C.L.  For the latter
mixing case, we determine $\xi_{af} = (0.32\pm0.16\pm0.12)\%$ or
$<0.91 \%$ at the 90\% C.L.  Figure~\ref{fig:f0mixing} b) shows our
results and upper limits compared to models ($q\bar{q}$, $q^2
\bar{q}^2$, $K\bar{K}$, and $q\bar{q} g$) and calculated values (SND,
KLOE, BNL, and CB)
based on $f_0$ and $a_0$ parameters.  References for models and
calculations may be found in Refs.~\cite{f0mixing} and
\cite{a0mixing}. Ours are the first direct measurements.

\section{\boldmath $X(1860)$ and $X(1835)$}
\begin{figure}[!htb]
\centerline{\psfig{file=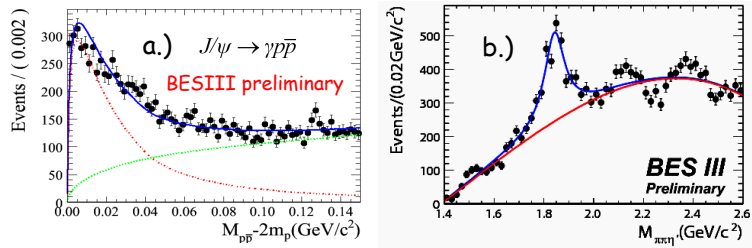,width=10.0cm}}
\vspace*{8pt}
\caption{a.) Threshold enhancement in $p\bar{p}$ mass in $J/\psi \to
  \gamma p\bar{p}$ b.)  $\eta' \pi^+\pi^-$ mass distribution in
  $J/\psi \to \gamma \eta' \pi^+\pi^-$ where $\eta' \to \pi^+\pi^-
  \eta$ and $\gamma \rho$.
  \label{fig:1860}}
\end{figure}

BESII observed a $p\bar{p}$ threshold enhancement in $J/\psi \to
\gamma p\bar{p}$.~\cite{ppbar} Fitted to a $S$-wave resonance, it gives
a mass below $p\bar{p}$ threshold of
$M(p\bar{p})=1859^{+3+5}_{-10-25}$ MeV/c$^2$.  BESIII has confirmed
this observation~\cite{b3ppbar} in $\psi' \to \pi^+\pi^- J/\psi, J/\psi
\to \gamma p\bar{p}$ with a fitted mass of $M(p\bar{p})=1861^{+6+7}_{-13-26}$
MeV/c$^2$ and directly in $J/\psi \to \gamma p\bar{p}$ with a fitted mass of
$M(p\bar{p})=1859\pm 0.8$ (preliminary and statistical error only),
shown in Fig.~\ref{fig:1860} a).  CLEOc also observes a $p\bar{p}$ threshold
enhancement in $\psi' \to \pi^+\pi^- J/\psi, J/\psi \to \gamma
p\bar{p}$ and obtains a consistent fitted mass.~\cite{cleopp}

BESII also observed a resonance at $M(\eta'
\pi^+\pi^-)=1833.7\pm6.1\pm2.7$ MeV/c$^2$ in $J/\psi \to \gamma \eta'
\pi^+\pi^-$ where $\eta' \to \pi^+\pi^- \eta$ and $\gamma
\rho$.~\cite{1835} BESIII also confirms this observation with a mass of
$M(\eta' \pi^+\pi^-)=1842.4\pm 2.8$ (preliminary and statistical error
only), as shown in Fig.~\ref{fig:1860} b).

\section{Summary and future prospects}

Many results, many preliminary, have been presented based on BESIII
samples of 106 M $\psi'$ and about 226 $J/\psi$ events.  Many more
results are to be expected in the future.  In addition, BESIII
acquired nearly $1 fb^{-1}$ of data at the $\psi(3770)$ resonance this
year, including approximately $75 pb^{-1}$ of scan data around the
$\psi(3770)$ peak.  Decays of the $\psi(3770)$ produce quantum
correlated $D \bar{D}$ pairs, which are ideal for mixing and CP
violation studies, as well measurements of absolute branching
fractions and studies of semi-leptonic decays. This sample allows
BESIII to begin their charm physics program.




\begin{thebibliography}{0}
\bibitem{CLEOcft} Values for $f_t$ are calculated from results in
  Ref.~\cite{CLEOcgV}.
\end{thebibliography}
\end{document}